\pdfoutput=1

\documentclass[11pt]{article}

\usepackage{acl}

\usepackage{times}
\usepackage{latexsym}

\usepackage[T1]{fontenc}

\usepackage[utf8]{inputenc}

\usepackage{microtype}

\usepackage{inconsolata}

\usepackage{graphicx}

\usepackage{subcaption}
\usepackage{listings}
\usepackage{lipsum}

\usepackage{pgfplots}
\pgfplotsset{compat=1.16}
\usepackage{comment}

\newcommand\siwarex{siwarex}

\title{A System and Benchmark for LLM-based Q\&A on Heterogeneous Data}

 \author{Achille Fokoue \and Srideepika Jayaraman \and Elham Khabiri \\
 \and {\bf Jeffrey O. Kephart} \and {\bf Yingjie Li} \and {\bf Dhruv Shah}  \and {\bf Youssef Drissi}\\
 \and {\bf Fenno F. Heath III} \and {\bf Anu Bhamidipaty} \and {\bf Fateh A. Tipu} \and {\bf Robert J.Baseman}  \\
         IBM Research}

\begin{document}
\maketitle
\begin{abstract}
In many industrial settings, users wish to ask questions whose answers may be found in structured data sources such as a spreadsheets, databases, APIs, or combinations thereof. Often, the user doesn't know how to identify or access the right data source. This problem is compounded even further if multiple (and potentially siloed) data sources must be assembled to derive the answer. Recently, various Text-to-SQL applications that leverage Large Language Models (LLMs) have addressed some of these problems by enabling users to ask questions in natural language. However, these applications remain impractical in realistic industrial settings because they fail to cope with the data source heterogeneity that typifies such environments. In this paper, we address heterogeneity by introducing the {\em \siwarex} platform, which enables seamless natural language access to both databases and APIs. To demonstrate the effectiveness of {\em \siwarex}, we extend the popular Spider dataset and benchmark by replacing some of its tables by data retrieval APIs. We find that {\em \siwarex} does a good job of coping with data source heterogeneity. Our modified Spider benchmark will soon be available to the research community.

\end{abstract}

\section{Introduction} 

In recent years, significant performance improvements in Large Language Models (LLMs) have driven the widespread adoption of natural language interfaces that access, retrieve, and reason over a variety of structured data sources  to answer questions. In enterprise settings, this paradigm shift promises to democratize access to critical data and analytics by users with limited knowledge of formal structured query languages (e.g., SQL to access databases) or API invocation. 

In practically any industrial environment, one finds a mixture of different types of structured data sources, including various flavors of SQL databases, NoSQL databases, and APIs. A master DB might contain records on sales, customers, or assets. Time series data might be available from an API such as the OSISoft PI web interface or an instance of the GE Digital Historian. Sales forecasting and other analytical capabilities may also be available via APIs.

Unfortunately, existing Q\&A systems do not support such heterogeneous environments. Some, like SQL tools used in the Langchain \cite{Langchain}, are designed with the assumption that all structured information is available from a single DB system, such as postgres. Others (e.g., ~\cite{patil2023gorilla}) assume that all data are obtained by accessing APIs. Unsurprisingly, existing benchmarks reinforce this unrealistic assumption of homogeneity. Multiple benchmarks~\cite{yu-etal-2018-spider,zhong2017seq2sql, li2023can} have been developed to assess the capability of LLMs to successfully convert natural language questions into SQL queries against a database (the text-to-SQL problem). In particular, all data that comprise the Spider~\cite{yu-etal-2018-spider} benchmark are stored in {\em SQLite} 
tables. Many other benchmarks~\cite{patil2023gorilla, li2023api} evaluate the ability of LLMs to invoke the right APIs to answer a user’s question. However, no existing benchmarks measure the efficacy of tools that cope with heterogeneous data sources.


The purpose of this paper is to introduce a system (\siwarex) and a benchmark that explicitly address this critical gap.  After a brief review of relevant literature in section~\ref{sec:relatedwork}, in section~\ref{sec:framework}, we introduce \textit{\siwarex}, a framework that supports question-answering across heterogeneous data sources consisting of a mixture of different types of databases and API calls. \textit{\siwarex} does so by exploiting:
\begin{enumerate}
    \item an LLM for the natural language understanding of the question and the generation of SQL statements given a unified relational representation of both available databases and APIs (where APIs are represented as \textit{virtual tables}),

    \item a Query Rewrite module that rewrites the LLM generated SQL queries by replacing mentions of \textit{virtual tables} with User-Defined Functions (UDFs) that invoke external APIs. By leveraging the UDF capability of modern database systems (e.g., postgres or DB2) to invoke external APIs, {\siwarex} places APIs on the same footing as database tables.
\end{enumerate}

Since access to any type of database (even NoSQL) can be achieved through API calls, {\siwarex} also has the potential to unify data sources across a spectrum of DB technologies.

We have already successfully deployed {\siwarex} in real industry applications in two domains: Oil\&Gas and Energy\&Utility. To demonstrate the effectiveness of {\siwarex} on publicly available data, we extend the popular Spider dataset and benchmark~\cite{yu-etal-2018-spider} to simulate the more realistic heterogeneous data environments that we encountered in our two industrial deployments. In particular, we replace some Spider tables with data retrieval APIs, as detailed in section~\ref{sec:dataset}. Then, in section~\ref{sec:evaluation}, we measure \textit{\siwarex}'s Q\&A accuracy as a function of the proportion of DB calls vs. API calls, showing that it does a reasonable job of coping with data source heterogeneity. To encourage further efforts to develop high-quality LLM-based Q\&A systems for industry, we are making this modified Spider benchmark available to the worldwide research community. We conclude with a summary and a brief discussion of some limitations of our work.

\section{Related work} 
\label{sec:relatedwork}
In this section, we overview the most relevant prior benchmarks and LLM-based Q\&A tools.

Two of the most popular pure Text-to-SQL benchmarks are Spider~\cite{yu-etal-2018-spider} and BIRD~\cite{li2023can}. The Spider benchmark is a text-to-sql dataset that contains a set of databases that each possess multiple tables and an associated set of natural language queries for which the correct SQL translation is known. The ground truth for a given natural language query is an SQL statement that appropriately retrieves and filters information from the relevant tables. An agent's benchmark score reflects how well its generated SQL statements match the ground truth averaged over the set of natural language queries. 

One popular pure API benchmark is APIBench~\cite{patil2023gorilla}, which has introduced a dataset consisting of HuggingFace, TorchHub, and TensorHub APIs. Another is API-Bank~\cite{li2023api}, a benchmark for tool-augmented LLMs that includes an evaluation system with API tools and tool-use dialogues to assess the capabilities of a given LLM in planning, retrieving, and calling APIs.

In contrast to these pure Text-to-SQL or API benchmarks, the benchmark we introduce in Section~\ref{sec:dataset} is the first to assess a system's ability to handle a mixture of DB accesses and API calls to answer natural language questions.


DAIL-SQL~\cite{gao2023texttosqlempoweredlargelanguage} is the Text-to-SQL method that gained the highest position on the Spider Leaderboard among opensource solutions (second place overall). The authors provided a comprehensive analysis of the performance of various open and closed source LLMs in various settings (pre-trained vs fine-tuned vs aligned and zero-shot vs multiple-shot).
Their experimental findings (their Table 3) motivated our use of the Mixtral 8x7b model~\cite{jiang2024mixtral}, on the grounds that it achieves nearly the best execution accuracy on zero-shot evaluation on the Spider benchmark (0.66 vs 0.68 for CodeLLAMA-34B, the best open-source model) and it has a more liberal license.


Gorilla~\cite{patil2023gorilla} is a LLaMA-based model for writing API calls by using self-instruct fine-tuning and retrieval to select from a large, overlapping, and changing set of APIs. We have chosen to employ it in our baseline experiments of Section~\ref{sec:evaluation} because it has performed very well on the APIBench benchmark.




\section{The \textit{\siwarex} Framework} 
\label{sec:framework}

This section describes the components of our
\textit{\siwarex} framework, which is built on top of a ReAct~\cite{yao2022react} agent implemented using the LangChain ReAct Agent\footnote{python.langchain.com/v0.1/docs/modules/agents/\\agent\_types/react/}. 



\textit{\siwarex} leverages two data source schema that can be provided by users or derived semi-automatically from domain metadata:
\begin{enumerate}
    \item The \textit{Abstract Schema}, in the form of an Entity-Relationship Diagram, provides a global view of the data source properties and  interrelationships in a format that is agnostic to whether the data source is a database table or an API.
    \item The \textit{API Mapping Schema} provides information necessary to invoke an API call, such as the URL, the method (POST, GET, etc.), and details of the input and output parameters.
\end{enumerate}

\begin{figure*}[hbtp]
\centering
   \includegraphics[scale=0.50]{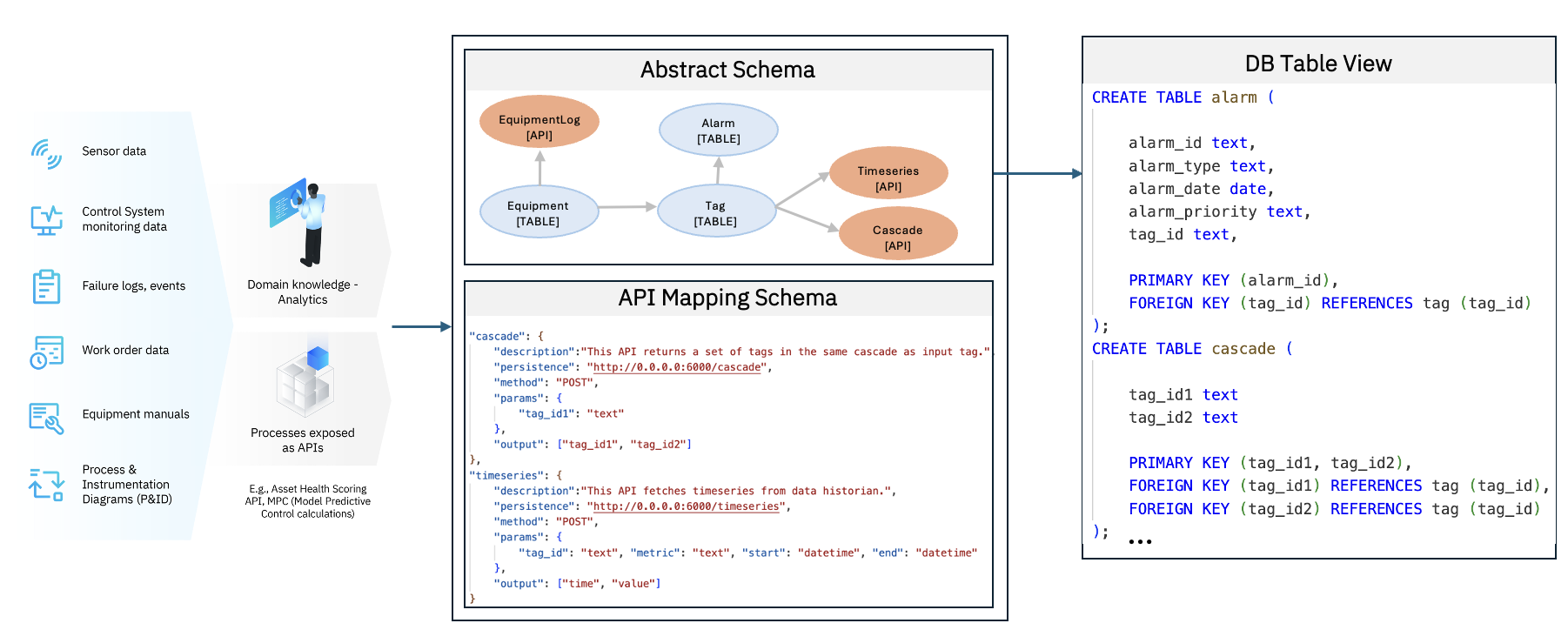}
    \caption{\textbf {Example of schemas and table view used by \siwarex.} The Abstract Schema and API Mapping Schema required by {\em \siwarex} can be provided manually or extracted from domain metadata. If a database schema is provided, the Abstract Schema can be extracted from it automatically; likewise the API Mapping Schema can be extracted automatically from an OpenAPI spec. For systems that mix DB access and API calls, the edges between API and DB nodes in the Abstract Schema may be augmented by a minimal amount of expert knowledge. Once the Abstract Schema is created, a relational schema (DB Table View) is generated from it automatically. The DB Table View represents all entities consistently as tables regardless of whether they are actually tables or APIs.}
    \label{fig:SchemaExtraction}
\end{figure*}

As explained in Figure~\ref{fig:SchemaExtraction}, the Abstract Schema and API Mapping Schema are generated either manually or via an automated process that leverages database schemas, OpenAPI specs, or other metadata. A deterministic offline process automatically converts the Abstract Schema into a relational schema that is used at runtime to generate SQL queries from NL questions. In that relational schema, tables corresponding to APIs (e.g., cascade) are \textit{virtual tables}, each of which is associated with a corresponding User Defined Function (UDF) that invokes the associated API by consulting details provided in the API Mapping.

\begin{figure*}[hbtp]
    \centering
   \includegraphics[scale=0.3]{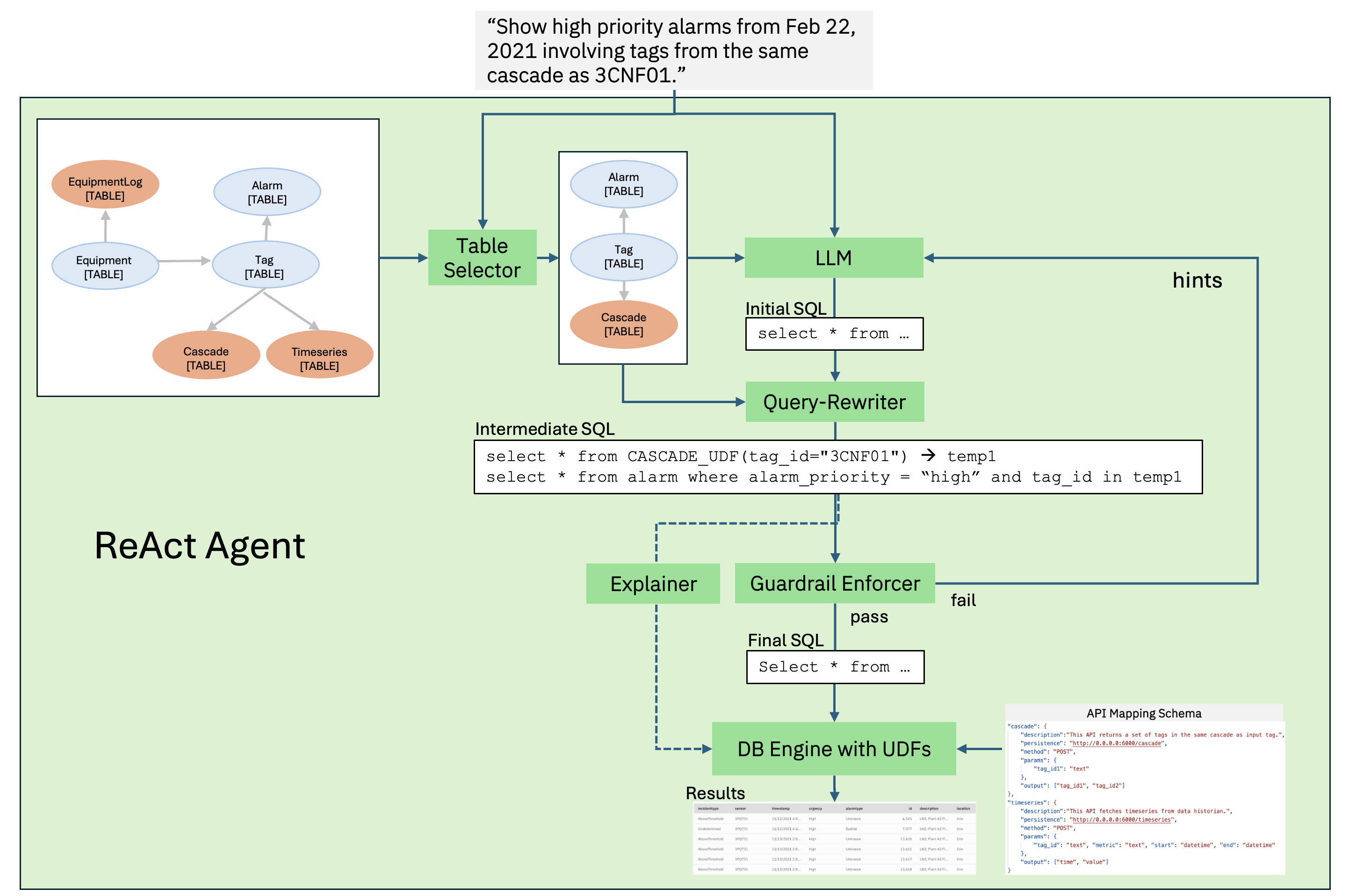}
    \caption{\textbf{Runtime view of the main {\em \siwarex} components} described in Section~\ref{sec:framework}.
    In this example, {\em \siwarex} produces an answer in response to ``Show high priority events from February 22 involving tags from the same cascade as 3CNF01.'' If the user opts to receive intermediate explanations of the system's thought process, the explainer routes intermediate results through the DB engine, which can then (like the final output) be rendered as text, tables, or other graphic formats.}
    \label{fig:runtime}
\end{figure*}

The \textit{\siwarex} framework comprises several components that are illustrated in Figure~\ref{fig:runtime}:

\begin{itemize}
    \item \textbf{ReAct Agent}~\cite{yao2022react} manages the step-by-step reasoning process using a specially-designed LLM prompt (see Appendix \ref{sec:appendixA}) in conjunction with 
        (a) tools that retrieve schema for specified tables, and (b) a \textbf{DB Engine} that executes intermediate or final generated SQL queries against the DB after they have been rewritten by the \textbf{Query Re-Writer} (see below) to use UDFs to execute APIs represented by virtual tables in the relational schema.
        
        \item \textbf{Table Selector} uses a Mixtral 8x7b~\cite{jiang2024mixtral} to examine the input question and identify the tables that are most likely to be required by the ReAct agent in order to produce an answer. Without such a selection process, the metadata could be too voluminous to be included in the prompt, particularly if the number of tables and/or columns is large.
        \item \textbf{Query Re-Writer} bridges the physical and logical representation of data entities by rewriting LLM generated SQL queries containing \textit{virtual tables} into executable SQL queries through the replacement of \textit{virtual tables} with their corresponding UDFs - including passing the right arguments (based on a static analysis of the generated SQL query)
    \item \textbf{Guardrail Enforcer} monitors the step-by-step reasoning process of the ReAct agent and applies a set of hints or rules that enable the ReAct Agent to detect and correct errors. Examples of such guardrail rules include:
    (1) entity names \& properties must be valid,
    (2) API signatures \& params must be valid,
    (3) column names used in sub-queries must exist within their associated table schema.
    If a violation is detected, a hint may be provided that enables the LLM to either repair the problem (e.g. amend the column or table name) or backtrack and try to generate a different query that does not violate any guardrails.
    \item \textbf{Explainer} enables the ReAct agent to explain its reasoning process to the user at each step. Multiple industry clients have told us that explainability is very important, but so is data privacy. A na\"ive implementation of explainability would require the LLM to execute intermediate queries and use their results as inputs to the next step --- thereby exposing the LLM to private data. To avoid this, we store intermediate results in temporary tables whose contents are not shared directly with the LLM. It is only necessary for the LLM to know whether or not a given sub-query successfully generated a temporary table; if so it simply uses that table's \emph{name} as a symbol that it can manipulate in subsequent steps.  
    

\end{itemize}

\section{New benchmark datasets} 
\label{sec:dataset}
In order to assess the ability of {\em \siwarex} or any other agent to cope with heterogeneous data sources, we need an appropriate benchmark. Since no such benchmark exists currently, we must create one. While in theory it might be possible to pool a large set of questions and answers adapted from real industrial Q\&A examples, such an approach would likely face severe practical and political obstacles. Instead, we have opted to modify the popular Spider benchmark~\cite{yu-etal-2018-spider}. 

To assess the ability of {\em \siwarex} or any other agent to translate utterances into an appropriate mixture of database (DB) and API calls, we extend the Spider benchmark 
by randomly replacing an adjustable percentage {\em ATTR} (the API to Table Ratio) of the DB tables with equivalent API calls. An {\em ATTR} of 0\% reduces to the traditional Spider benchmark, while an {\em ATTR} of 100\% represents the opposite extreme in which all data required to answer a question must be accessed via APIs. Once APIs are generated for a given {\em ATTR}, the replaced tables are deleted from the original database.

The generated APIs serve as proxies for the replaced tables, thereby converting a database retrieval task into an API call with the appropriate parameters. For example, if a certain database query involves filtering on a table based on 2 columns, this would translate into calling the corresponding API with a request consisting of 2 parameters being set to the value on which they need to be filtered. 

Many natural language queries in Spider require combining information from multiple tables. For intermediate values of {\em ATTR}, a correct answer to a particular query may require combining at least one DB call with at least one API call.

We illustrate our revised benchmark with an example from the Spider dataset involving the database $museum\_visit$, which contains 3 tables: $museum$, $visitor$ and $visit$. Consider the table $museum$, which has the following SQL definition:
\begin{lstlisting}
CREATE TABLE "museum" ("Museum_ID" int,
"Name" text,"Num_of_Staff" int,
"Open_Year" text, PRIMARY KEY ("Museum_ID")
\end{lstlisting}

\noindent If the $museum$ table is replaced by an API, it can be represented by the swagger definition detailed in Appendix B. 

First, consider the simple natural language question: ``What are the opening year and staff number of the museum named Plaza Museum?''. In the original benchmark, the correct approach is to convert this to the SQL statement:  \textit{"SELECT Num\_of\_Staff , Open\_Year FROM museum WHERE name = 'Plaza Museum'"}. However, with the DB table $museum$ replaced by the API $/museum$, the correct approach is to call the $/museum$ API with the parameter $name=``Plaza Museum''$.

Now consider a more complex question: \textit{"What are the id, name and membership level of visitors who have spent the most money in total in all museum tickets?"}. Whereas in the traditional Spider benchmark this would entail joining the $museum$ and $visit$ tables, in our extended benchmark this requires combining results of a DB call to the $visit$ table with results of an API call to $/museum$.

\section{Evaluation} 
\label{sec:evaluation}
To evaluate {\em \siwarex}, we compare its performance with that of a simple baseline consisting of a ReAct agent that does not include most of the {\em \siwarex} components introduced in Section~\ref{sec:framework}. 


The Baseline agent uses two state-of-the-art tools:
\begin{enumerate}
    \item Gorilla~\cite{patil2023gorilla}, currently the top open-source API-invoking system on the Berkeley Function-Calling Leaderboard\footnote{gorilla.cs.berkeley.edu/leaderboard.html}; and
    \item SQLDatabaseToolkit, a tool for interacting with SQL databases, from LangChain community\footnote{https://api.python.langchain.com/en/latest/\\community\_api\_reference.html} configured to use Mixtral 8x7~\cite{jiang2024mixtral} as its LLM.
\end{enumerate}

\noindent It decomposes an input question into more basic natural language questions, each of which can be answered by retrieving the required information from a single data modality. Basic questions requiring information from DB tables are routed to the SQLDatabaseToolkit,
while those requiring information available from APIs are sent to the Gorilla tool. As shown in Figure~\ref{fig:baseline_prompt}, the ReAct prompt explicitly includes the list of available APIs and DB tables, thereby enabling the Baseline agent to properly decompose the input questions into single-modality basic questions and route those basic questions to the right tool.

\begin{figure}
\begin{tikzpicture}[scale=0.80]
\begin{axis}[
    xlabel=$ATTR$,
    ylabel=$Execution$ $Accuracy$,
    xmin=0, xmax=50,
    ymin=0, ymax=1,
    xtick={10,20,30,40,50},
    xticklabels={20\%,40\%,60\%,80\%,100\%},   
    ytick={0.0,0.2,...,1.0},
    title=Total Difficulties
            ]
\addplot[smooth,color=blue,mark=*] plot coordinates {
    (0,0.68)
    (10,0.62)
    (20,0.32)
    (30,0.20)
    (40,0.15)
    (50,0.08)
    };
\addlegendentry{Baseline}

\addplot[smooth,mark=x,red] plot coordinates {
    (0,0.66)
    (10,0.66)
    (20,0.64)
    (30,0.61)
    (40,0.58)
    (50,0.56)
};
\addlegendentry{\siwarex}

\end{axis}
\end{tikzpicture}
\caption{Accuracy comparison overall questions, regardless of difficulty.}
\label{fig:allDifficulty}
\end{figure}
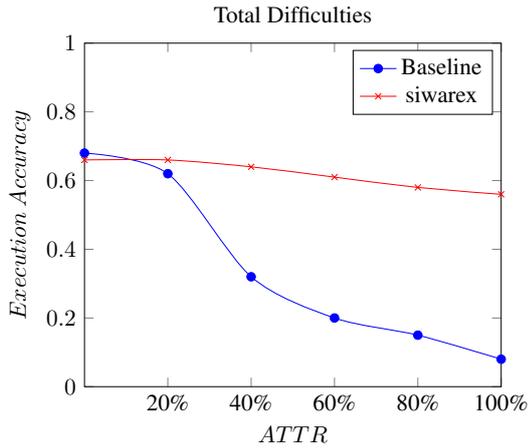

The evaluation metric is the execution accuracy measured by comparing the results produced by our system (or the baseline system) against those produced by the evaluation of the gold standard spider sql query on the original spider db. We use the sophisticated comparison approach that was introduced by~\cite{zhong2020}. For {\em \siwarex} and the Baseline, the execution accuracy aggregated over all levels of difficulty is plotted in Figure~\ref{fig:allDifficulty} as a function of {\em ATTR} (the API to Table Ratio). 

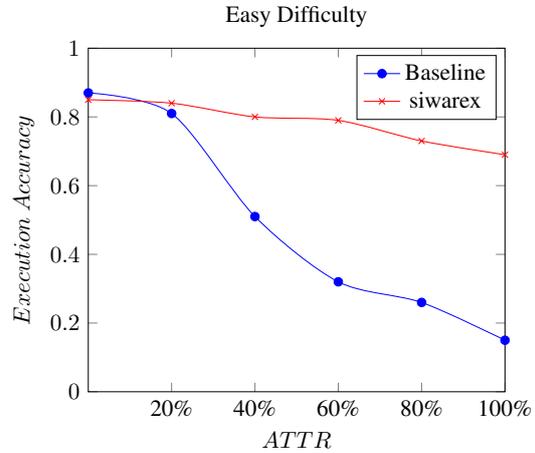
\begin{figure}
\begin{tikzpicture}[scale=0.80]
\begin{axis}[
    xlabel=$ATTR$,
    ylabel=$Execution$ $Accuracy$,
    xmin=0, xmax=50,
    ymin=0, ymax=1,
    xtick={10,20,30,40,50},
    xticklabels={20\%,40\%,60\%,80\%,100\%},   
    ytick={0.0,0.2,...,1.0},
    title=Easy Difficulty
            ]
\addplot[smooth,color=blue,mark=*] plot coordinates {
    (0,0.87)
    (10,0.81)
    (20,0.51)
    (30,0.32)
    (40,0.26)
    (50,0.15)
    };
\addlegendentry{Baseline}

\addplot[smooth,mark=x,red] plot coordinates {
    (0,0.85)
    (10,0.84)
    (20,0.80)
    (30,0.79)
    (40,0.73)
    (50,0.69)
};
\addlegendentry{\siwarex}

\end{axis}
\end{tikzpicture}
\caption{Accuracy comparison for questions of easy difficulty.}
\label{fig:easyDifficulty}
\end{figure}

\begin{figure}
\begin{tikzpicture}[scale=0.80]
\begin{axis}[
    xlabel=$ATTR$,
    ylabel=$Execution$ $Accuracy$,
    xmin=0, xmax=50,
    ymin=0, ymax=1,
    xtick={10,20,30,40,50},
    xticklabels={20\%,40\%,60\%,80\%,100\%},   
    ytick={0.0,0.2,...,1.0},
    title=Extra Hard Difficulty
            ]
\addplot[smooth,mark=*,blue] plot coordinates {
    (0,0.33)
    (10,0.25)
    (20,0.10)
    (30,0.06)
    (40,0.03)
    (50,0.03)
};
\addlegendentry{Baseline}

\addplot[smooth,color=red,mark=x] plot coordinates {
    (0,0.34)
    (10,0.33)
    (20,0.34)
    (30,0.28)
    (40,0.29)
    (50,0.27)

    };
\addlegendentry{\siwarex}

\end{axis}
\end{tikzpicture}
\caption{Accuracy comparison for questions of extra hard difficulty.}
\label{fig:extraHardDifficulty}
\end{figure}
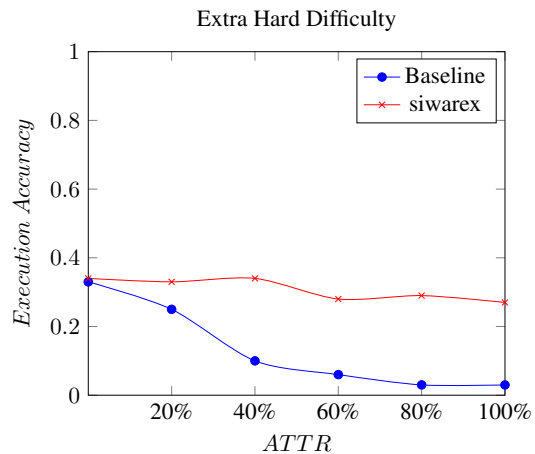

Without any APIs (at 0\%), both {\siwarex} and the baseline system perform very close to the state-of-the-art open-source LLM (CodeLLAMA-34B) on Zero-shot evaluation (see Table 3 in ~\cite{zhong2020} which reports 0.68 accuracy for CodeLLAMA-34B). However, as the proportion of API calls increases, the performance of the baseline system deteriorates significantly whereas the performance of {\siwarex} degrades only modestly. Figs.~\ref{fig:easyDifficulty} and~\ref{fig:extraHardDifficulty}, which plot execution accuracy over ``easy'' and ``extra hard'' questions respectively, tell a similar story. As expected, the accuracy for easy and extra hard questions are above and below that of the question set as a whole, and those for ``medium'' and ``hard'' questions are bracketed between these two extremes (Figures~\ref{fig:mediumDifficulty} and~\ref{fig:hardDifficulty}.)

A careful error analysis suggests that several key issues account for much of the baseline system's poor performance:
\begin{enumerate}
    \item \textbf{Sequencing}. State-of-the-system function calling LLM based systems are not trained, fine-tuned or optimized to perform complex API sequencing, merging, and aggregation tasks. They tend to perform relatively well on questions whose answers require performing a single API call. However, on our benchmark with an {\em ATTR} of 100\%, even the easier Spider queries (e.g., "What is the total number of singers?") typically require invoking multiple APIs and sequencing them properly (e.g., invoke getAllSingers() and then invoke getSize() on the previous result).
    \item \textbf{Routing}. On datasets with a mixture of APIs and DB tables, the master LLM (Mixtral 8x7) often decomposes the problem correctly, but then it fails to properly route the decomposed questions to the tool (db or API tool) that has the right information to answer it.  
    \item \textbf{Hallucination}. Even when the right API is selected for invocation, the arguments for its invocation are often hallucinated values.
\end{enumerate}
{\siwarex} avoids all the above issues by providing to an LLM a single relational view that removes all the complexity of dealing with multiple heterogeneous sources. To the LLM, everything appears to be relational, and thus it can leverage its Text2SQL capability to generate a SQL query for each NL question. The Query Rewriter is then responsible to inject API invocations through UDFs with the proper arguments (inferred from a deterministic analysis of the SQL query).

\section{Conclusion} 
In this paper, we introduce both {\siwarex}, a novel system for LLM-based Q\&A over heterogeneous data source, and a new benchmark (which we plan to make publicly available) that simulates the common heterogeneous data environments encountered in various industries. Our experimental evaluation shows that {\siwarex} performs remarkably well at different levels of data heterogeneity compared to a baseline made of two independent open-source state-of-the-art systems that access databases and APIs, respectively.

\section{Limitations}

One limitation that we are eager to address in future work is that our benchmark's evaluation metric only considers the execution {\em accuracy} of the final results. Especially since we wish to produce a benchmark that meaningfully captures practical issues that arise in industrial settings, it is incumbent on us to augment this metric with an execution {\em performance} metric (i.e. efficiency or speed) like that introduced by BIRD~\cite{li2023can}.

Additionally, the insights we derived in Section~\ref{sec:evaluation} should now be factored into a new version of {\em \siwarex} that can be tested with our new benchmark (with the extra performance metric included).

Finally, the current work only considers a limited form of data heterogeneity in which DB accesses are replaced with API calls. While this is a natural and convenient way to extend the Spider benchmark, we can further improve the realism of our benchmark by introducing greater diversity in the type of databases being used (for example, mixing in other types of SQL databases along with some NoSQL databases) and introducing APIs that are not just replacements for database calls but perform calculations (e.g. statistical operations) or analytics (e.g. timeseries correlations). 

\bibliography{custom}
\pagebreak

\appendix
\section*{Appendix A}
\label{sec:appendixA}
\begin{figure}[htbp]
\centering
\begin{tabular}{c}
\begin{lstlisting}
You are an agent designed to answer questions.
Here is the list of all available tools along
with their descriptions.

entity_tool: This tool answers natural language
questions about the following entity types: 
{table_names}. The input is a question in English. 
The input CANNOT be in JSON format, it has to be 
in English. The output is the answer to the 
question as a list of multiple entities in JSON 
format.

api_tool: This tool answers natural language 
questions that require access to the following apis: 
{apis}. The input is a question in English. The 
input CANNOT be in JSON format, it has to be in 
English. The output is the answer to the question in 
JSON format.

Use the following format:

Question: the input question you must answer
Thought: you should always think about what to do
Action: the action to take, should be one of 
[entity_tool, api_tool]
Action Input: the input to the action
Observation: the result of the action
... (this Thought/Action/Action Input/Observation 
can repeat N times)
Thought: I now know the final answer
Final Answer: the final answer to the original input
question. Show the final answer as JSON in a single
line. Begin!

Question: {input}
{agent_scratchpad}

\end{lstlisting}
\end{tabular}
\caption{Baseline ReAct Prompt.}
\label{fig:baseline_prompt}
\end{figure}

\section*{Appendix B}
\label{sec:appendixB}
\begin{lstlisting}[language=Python, caption=ReAct Agent prompt 
(extension of the base sql ReAct LangChain agent)]

customized_SQL_PREFIX = """
    You are an agent designed to interact with a SQL 
    database. Given an input question, create a 
    syntactically correct {dialect} query to run. 
    Never query for all the columns from a specific 
    table, only ask for the relevant columns given 
    the question. You have access to tools for 
    interacting with the database. Only use the below 
    tools. Only use the information returned by the 
    below tools to construct your final answer. If you 
    get an error while executing a query, rewrite the 
    query and try again.

    DO NOT make any DML statements (INSERT, UPDATE, 
    DELETE, DROP, CREATE etc.) to the database.

    VERY IMPORTANT: Before using any of the following 
    tables [{materialized_tables}], you must first 
    invoke the appropriate tool that populates them.

    If the question does not seem related to the 
    database, just return "I don't know" as the answer.

    Here is the list of all available tools along with 
    their descriptions.

    {tool_descriptions}
    """

customized_FORMAT_INSTRUCTIONS = """
   Use the following format:

   Question: the input question you must answer
   Thought: you should always think about what to do. 
   Very brief description of your thought
   Action: the action to take, should be one of 
   [{tool_names}]
   Action Input: the input to the action
   Observation: the result of the action
   ... (this Thought/Action/Action Input/Observation 
   can repeat N times)
   Thought: you should perform your final action
   Action: The final action to take to execute the 
   final query. This final action should be 
   sql_db_query.
   Action Input: the input to the action. The input 
   to this final action should be the final query 
   whose execution returns the final answer.
   Observation: the result of the action
   Thought: I now know the final answer
   Final Answer: The sql query whose evaluation 
   would provide the final answer"""

customized_SUFFIX= """
    {user_input}
    {model_answer_prefix}
    {agent_scratchpad}"""

prompt = customized_SQL_PREFIX 
        + customized_FORMAT_INSTRUCTIONS 
        + customized_SUFFIX 
\end{lstlisting}

\section*{Appendix C}
\label{sec:appendixC}



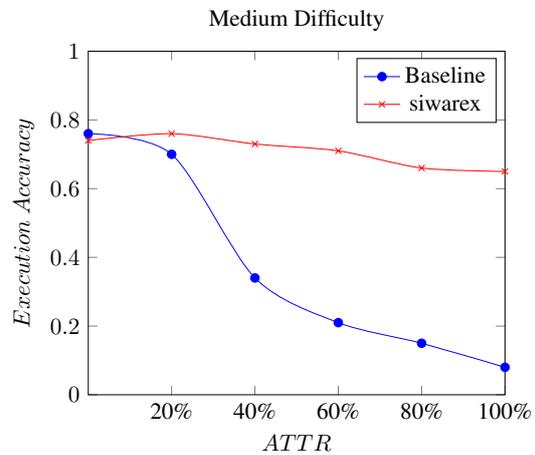
\begin{figure}[htbp]
\begin{tikzpicture}[scale=0.80]
\begin{axis}[
    xlabel=$ATTR$,
    ylabel=$Execution$ $Accuracy$,
    xmin=0, xmax=50,
    ymin=0, ymax=1,
    xtick={10,20,30,40,50},
    xticklabels={20\%,40\%,60\%,80\%,100\%},   
    ytick={0.0,0.2,...,1.0},
    title=Medium Difficulty
            ]
\addplot[smooth,mark=*,blue] plot coordinates {
    (0,0.76)
    (10,0.70)
    (20,0.34)
    (30,0.21)
    (40,0.15)
    (50,0.08)
};
\addlegendentry{Baseline}

\addplot[smooth,color=red,mark=x] plot coordinates {
    (0,0.74) 
    (10,0.76)
    (20,0.73)
    (30,0.71)
    (40,0.66)
    (50,0.65)
    };
\addlegendentry{\siwarex}

\end{axis}
\end{tikzpicture}
\caption{Accuracy comparison for questions of medium difficulty.}
\label{fig:mediumDifficulty}
\end{figure}

\begin{figure}[htbp]
\begin{tikzpicture}[scale=0.80]
\begin{axis}[
    xlabel=$ATTR$,
    ylabel=$Execution$ $Accuracy$,
    xmin=0, xmax=50,
    ymin=0, ymax=1,
    xtick={10,20,30,40,50},
    xticklabels={20\%,40\%,60\%,80\%,100\%},   
    ytick={0.0,0.2,...,1.0},
    title=Hard Difficulty
            ]
\addplot[smooth,mark=*,blue] plot coordinates {
    (0,0.50)
    (10,0.45)
    (20,0.17)
    (30,0.13)
    (40,0.08)
    (50,0.05)
};
\addlegendentry{Baseline}

\addplot[smooth,color=red,mark=x] plot coordinates {
    (0,0.48)
    (10,0.50)
    (20,0.46)
    (30,0.44)
    (40,0.45)
    (50,0.44)

    };
\addlegendentry{\siwarex}

\end{axis}
\end{tikzpicture}
\caption{Accuracy comparison for questions of hard difficulty.}
\label{fig:hardDifficulty}
\end{figure}
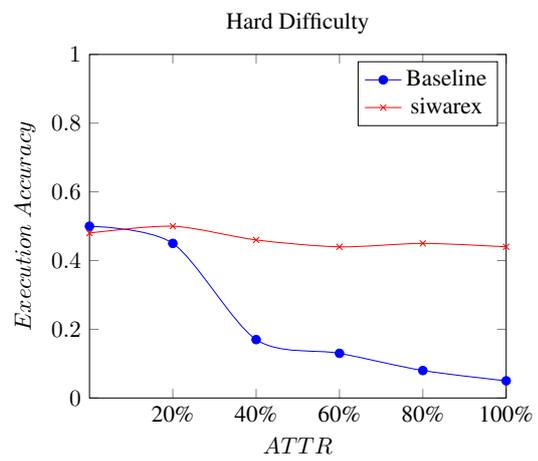





\end{document}